\documentclass[superscriptaddress,aps,preprintnumbers,showpacs,prd,nofootinbib,reprint]{revtex4-1}

\usepackage{graphicx} 
\usepackage{amsmath,amssymb,amsthm,slashed,bm}
\usepackage[colorlinks,allcolors=blue]{hyperref}

\usepackage{bm,epstopdf,natbib,hyperref,color,verbatim,multirow,bm,tikz,xcolor}
\hypersetup{colorlinks=true,urlcolor=blue,citecolor=blue,linkcolor=blue,menucolor=blue,anchorcolor=blue,filecolor=blue}
\everymath{\displaystyle}
\definecolor{lime}{HTML}{A6CE39}
\DeclareRobustCommand{\orcidicon}{
	\begin{tikzpicture}
	\draw[lime, fill=lime] (0,0) 
	circle [radius=0.16] 
	node[white] {{\fontfamily{qag}\selectfont \tiny ID}};
	\draw[white, fill=white] (-0.0625,0.095) 
	circle [radius=0.007];
	\end{tikzpicture}
	\hspace{-3mm}
}
\foreach \x in {A, ..., Z}{\expandafter\xdef\csname orcid\x\endcsname{\noexpand\href{https://orcid.org/\csname orcidauthor\x\endcsname}
			{\noexpand\orcidicon}}
}

\begin{document}


\preprint{TU-1300}

\title{
Number Theory in Quantum Physics: Minicharged Particles and the Prouhet-Tarry-Escott Problem
}

\author{
Junseok Lee\hspace{-1mm}\orcidA{}
}
\email{lee.junseok.p4@dc.tohoku.ac.jp}
\affiliation{Department of Physics, Tohoku University, 
Sendai, Miyagi 980-8578, Japan}
\author{
Fuminobu Takahashi\hspace{-1mm}\orcidB{}
}
\email{fumi@tohoku.ac.jp}
\affiliation{Department of Physics, Tohoku University, 
Sendai, Miyagi 980-8578, Japan} 
\affiliation{Kavli IPMU (WPI), UTIAS, University of Tokyo, Kashiwa 277-8583, Japan}
\author{
Yu-Dai Tsai\hspace{-1mm}\orcidC{}
}
\email{yu-dai.tsai@manchester.ac.uk,yudaitsai.academic@gmail.com}
\affiliation{The University of Manchester, Manchester M13 9PL, United Kingdom}
\affiliation{Los Alamos National Laboratory (LANL), Los Alamos, NM 87545, USA}

\begin{abstract}
In quantum gauge theories, anomaly cancellation severely restricts the allowed patterns of chiral charges. Here we show that, in a phenomenologically motivated framework for light minicharged particles, the anomaly cancellation conditions are equivalent to the degree $k=3$ Prouhet-Tarry-Escott problem in number theory. This correspondence immediately implies that the hidden sector must contain at least four minicharged states. For constructions based on minimal ideal solutions, the mass spectrum generically exhibits a near-degenerate doublet structure, so that the discovery of one minicharged particle would point to a partner state with the same minicharge and a nearby mass. Our results uncover an unexpected link between quantum consistency and number theory, with direct implications for model building and future searches.
\end{abstract}

\maketitle
\flushbottom

\vspace{1cm}

{\bf Introduction.--}
Galileo famously remarked that the universe is written in the language of mathematics~\cite{Galilei:1623,Galilei:1957}.
From differential geometry in general relativity to the deep interplay between string theory and mathematics, this theme has repeatedly guided progress in fundamental physics.
In quantum gauge theories, this connection becomes especially sharp because discrete charge assignments are constrained by anomaly cancellation.
Here, we show that a phenomenologically motivated framework for light minicharged particles (mCPs) provides a concrete realization of this connection between fundamental physics and mathematics. Specifically, the anomaly cancellation conditions are equivalent to
a classic question in number theory, the degree $k=3$ Prouhet-Tarry-Escott (PTE) problem~\cite{Wright:1959Prouhet,Borwein2002,shuwen2025survey}.
The PTE problem asks for two lists of integers, allowing repetitions, whose power sums agree up to order $k$.

mCPs naturally arise in well-motivated extensions of the Standard Model (SM).
A simple realization introduces a hidden Abelian gauge symmetry $U(1)_{\rm H}$ that kinetically mixes with hypercharge, such that matter charged under $U(1)_{\rm H}$ acquires an effective minicharge under electromagnetism~\cite{Holdom:1985ag}.
Such scenarios have significant theoretical and cosmological implications~\cite{Dirac:1931kp,Pati:1973uk,Wen:1985qj,Gan:2023jbs,Boddy:2024vgt} and have become well-established benchmarks for physics beyond the SM. They are actively probed in laboratory searches~\cite{Prinz:1998ua,Magill:2018tbb,Kelly:2018brz,Harnik:2019zee,ArgoNeuT:2019ckq,Berlin:2019uco,Foroughi-Abari:2020qar,Plestid:2020kdm,milliQan:2021lne,Kling:2022ykt,SENSEI:2023gie,Oscura:2023qch,Tsai:2024wdh,Alcott:2025rxn,Bailloeul:2025fde,Berlin:2025hjs}, and are constrained by astrophysical observations~\cite{Mohapatra:1990vq,Davidson:1993sj,Chang:2018rso} and cosmological measurements~\cite{Davidson:2000hf,Dolgov:2013una,Vogel:2013raa}.
See Refs.~\cite{Jaeckel:2010ni,Arias:2012az,deMontigny:2023qft} for reviews.

For mCPs to be accessible to experimental searches, their masses must be sufficiently small.
However, in much of the literature, the mCP mass is treated as a free parameter introduced by hand in the Lagrangian.
In this Letter, we instead consider a framework in which a chiral symmetry protects the small mCP mass.
We introduce two Abelian symmetries with distinct roles: an unbroken $U(1)_{\rm H}$ accounts for the minicharge, while a chiral $U(1)_{\rm X}$ controls the mass.
The $U(1)_{\rm H}$ charges are vector-like on each massive Dirac pair, whereas the chiral $U(1)_{\rm X}$ forbids a bare mass term for the mCP fermions.
A small mass is then generated when $U(1)_{\rm X}$ is spontaneously broken.

For this setup to define a consistent quantum theory, all gauge anomalies must cancel.
Anomaly-free charge assignments under a $U(1)$ gauge symmetry have been studied in the literature, including in dark-matter motivated contexts~\cite{Nakayama:2011dj}, and explicit construction methods are known~\cite{Batra:2005rh,Costa:2019zzy}. 
Our central result is that, once we impose a phenomenologically motivated $U(1)_{\rm H}\times U(1)_{\rm X}$ structure for light mCPs---with $U(1)_{\rm H}$ unbroken and vector-like, and $U(1)_{\rm X}$ chiral and spontaneously broken---the anomaly-free charge-assignment problem becomes equivalent to the PTE problem in number theory.

This equivalence has immediate phenomenological consequences: the hidden sector must contain at least {\it four} mCP mass eigenstates. In addition, our scan of solutions suggests that minimal four-state realizations typically exhibit a near-degenerate doublet structure, so that discovering one mCP would generically imply a partner with the same minicharge and a nearby mass. This highlights the importance of experiments that can probe the mass spectrum, and it calls for a careful interpretation of cosmological limits in multi-state chiral sectors. Furthermore, in our setup, the massive $U(1)_{\rm X}$ gauge boson can open additional production and decay channels, potentially changing the optimal search strategy.

{\bf Setup.--}
We consider fermions charged under a hidden $U(1)_{\rm H}$ gauge symmetry with the gauge field $A'_\mu$.
The hidden photon can kinetically mix with the SM hypercharge:
\begin{align}
\mathcal{L} \supset - \frac{1}{4} B_{\mu\nu} B^{\mu\nu} - \frac{1}{4} F'_{\mu\nu} F'^{\mu\nu} - \frac{\chi}{2} B_{\mu\nu} F'^{\mu\nu},
\end{align}
where $B_{\mu\nu}$ and $F'_{\mu\nu}$ are the field strengths of the SM hypercharge and the hidden photon, respectively, and $\chi$ is the kinetic mixing parameter.
For $|\chi|\ll 1$, one may diagonalize the kinetic terms by the field redefinition $A'_\mu \rightarrow A'_\mu - \chi B_\mu$ to leading order of $\chi$.
For a Dirac fermion $\Psi$ charged under $U(1)_{\rm H}$, we have
\begin{align}
\mathcal{L} \supset \bar{\Psi} (i \slashed{\partial} - g_{\rm H} q_{\rm H}\slashed{A}') \Psi - m \bar{\Psi} \Psi,
\end{align}
where $g_{\rm H}$ is the $U(1)_{\rm H}$ gauge coupling,  $q_{\rm H}$ is the $U(1)_{\rm H}$ charge of $\Psi$ and $m$ is its mass.
As a result of the diagonalization,  $\Psi$ acquires an effective minicharge, $-\chi g_{\rm H} q_{\rm H}$.

We note that, in much of the literature, $m$ is treated as a free parameter.
In the absence of a protective symmetry, however, there is no reason to expect $m$ to lie far below the cutoff scale $M_c$ of the theory (e.g., the GUT or Planck scale). It would therefore not be expected to be light enough for terrestrial searches or for cosmological and astrophysical probes.

A simple way to obtain light mCP masses is to forbid bare mass terms by a symmetry and generate small but nonzero masses through its breaking.
We implement this idea by introducing an additional gauged chiral $U(1)_{\rm X}$ symmetry.
Concretely, we consider $U(1)_{\rm H} \times U(1)_{\rm X}$, where $U(1)_{\rm H}$ remains unbroken and acts vector-like, while $U(1)_{\rm X}$ is spontaneously broken at low energies and acts chirally on the mCP fermions.

We introduce $2n$ left-handed Weyl fermions $\psi_i$ with  charges $(q_{{\rm H},i},q_{{\rm X},i})$ under $U(1)_{\rm H} \times U(1)_{\rm X}$.
To break $U(1)_{\rm X}$, we add a dark Higgs $\phi$ with charges $(0,1)$ and vacuum expectation value (VEV) $v_d$.
To ensure a vector-like charge assignment under $U(1)_{\rm H}$, we take $q_{{\rm H},i} = +1$ for $i = 1, \dots, n$, and $q_{{\rm H},i} = -1$ for $i = n+1, \dots, 2n$.
We assume that no gauge invariant bilinear $\psi_i\psi_j$ exists between the two $U(1)_{\rm H}$ sectors, i.e.
$q_{{\rm X},i}+q_{{\rm X},j}\neq 0$ for any $i=1,\dots,n$ and $j=n+1,\dots,2n$, so the $U(1)_{\rm X}$ charge assignment is chiral and forbids vector-like mass terms.\footnote{$\psi_i\psi_j$ denotes the Lorentz invariant bilinear $\psi_i^T(i\sigma^2)\psi_j$.}
As a result, bare mass terms are forbidden by $U(1)_{\rm X}$, while masses arise from higher-dimensional operators.%

For $i=1,\dots,n$ and $j=n+1,\dots,2n$, the leading operators take the form
\begin{align}
    y_{ij} M_c \left(\frac{\phi^*}{M_c} \right)^{q_{{\rm X},i} + q_{{\rm X},j}} \psi_i \psi_j + {\rm h.c.}
\end{align}
for $q_{{\rm X},i} + q_{{\rm X},j} > 0$ and
\begin{align}
    y_{ij} M_c \left(\frac{\phi}{M_c} \right)^{-(q_{{\rm X},i} + q_{{\rm X},j})} \psi_i \psi_j + {\rm h.c.}
\end{align}
for $q_{{\rm X},i} + q_{{\rm X},j} < 0$, respectively.
Here $M_c$ is a cutoff scale and $y_{ij}$ are $\mathcal{O}(1)$ complex coefficients.

Once the dark Higgs $\phi$ develops a nonzero VEV $v_d$, the fermions acquire suppressed masses.
Defining the small parameter $\epsilon < 1$ by
\begin{align}
    \epsilon \equiv \frac{v_d}{M_c},
\end{align}
the $n\times n$ mass matrix elements scale as powers of $\epsilon$, with exponents fixed by the $U(1)_{\rm X}$ charges.
The resulting mass spectrum therefore depends sensitively on the charge assignment.
Determining consistent charge assignments is nontrivial, since they must satisfy anomaly cancellation conditions; we study these constraints in the next section.

The breaking of $U(1)_{\rm X}$ also yields a massive $U(1)_{\rm X}$ gauge boson and a massive dark Higgs.
For the parameter range of interest, their masses are set by $v_d$ and are typically well above the mCP masses.

{\bf Anomaly Cancellation Conditions and the PTE problem.--}
We need to find a set of charge assignments ${q_{{\rm X},i}}$ that satisfy the anomaly cancellation conditions.
For convenience, let us rewrite $a_i = q_{{\rm X},i}$ for $i = 1,\dots, n$ and $b_j = -q_{{\rm X},j}$ for $j = n+1, \dots, 2n$.
Then the anomaly cancellation conditions read
\begin{itemize}
    \setlength{\itemsep}{0cm}
    \setlength{\parskip}{0cm}
    \item $U(1)_{\rm X}$\,-\,graviton\,-\,graviton \\
    \begin{align}
        \label{eq: U(1)X-g-g}
        \sum_{i=1}^n a_i = \sum_{i=1}^n b_i
    \end{align}
    \item $(U(1)_{\rm X})^3$ \\
    \begin{align}
        \label{eq: U(1)X3}
        \sum_{i=1}^n a_i^3 = \sum_{i=1}^n b_i^3
    \end{align}
    \item $(U(1)_{\rm X})^2$\,-\,$U(1)_{\rm H}$ \\
    \begin{align}
        \label{eq: U(1)X2-U(1)H}
        \sum_{i=1}^n a_i^2 = \sum_{i=1}^n b_i^2
    \end{align}
\end{itemize}
Note that the anomaly cancellation condition for $U(1)_{\rm X}$-$(U(1)_{\rm H})^2$ is the same as Eq.~(\ref{eq: U(1)X-g-g}), and that of $(U(1)_{\rm H})^3$ is trivial due to the vector-like nature of $U(1)_{\rm H}$.

Finding the rational or integer solutions to Eqs.~(\ref{eq: U(1)X-g-g}), (\ref{eq: U(1)X3}), and (\ref{eq: U(1)X2-U(1)H}) is equivalent to the PTE problem  with degree $k = 3$ (see Refs.~\cite{Borwein2002,shuwen2025survey} for reviews).
The PTE problem asks for two distinct multisets of integers, $A = \{a_1, \dots, a_n\}$ and $B = \{b_1, \dots, b_n\}$, such that
\begin{align}
    a_1^\ell + \dots + a_n^\ell = b_1^\ell + \dots + b_n^\ell
\end{align}
for $\ell = 1, \dots, k.$
A solution is denoted by $A =_k B$, where $n$ is its size and $k$ is its degree.
The most nontrivial case of the PTE problem is when $n$ takes its smallest value, $n = k+1$, and in this case the solution is called {\it ideal}. We focus on minimal ideal solutions, and we discuss non-ideal solutions in the Supplemental Material.

An affine transformation, $A'=\{c a_i + d\}$ and $B'=\{c b_i + d\}$, maps a solution to another solution, where $c$ and $d$ are rational numbers with $c\neq 0$.
Solutions generated by the affine transformation are considered equivalent in the context of the PTE problem.
In our setup, however, they are physically distinct because they modify both the mass suppression and the coupling strength to the hidden photon.
Also, we always exclude the case where $A$ and $B$ share a common element, as it merely gives a trivial extension.

As revealed later, physically interesting solutions for this problem are the cases satisfying $n=2m$ and
\begin{align}
    \begin{split}
        A&=\{\pm a_1,\ldots,\pm a_m\}\equiv\{a_1,-a_1,\ldots,a_m,-a_m\},\\
        B&=\{\pm b_1,\ldots,\pm b_m\}\equiv\{b_1,-b_1,\ldots,b_m,-b_m\},
    \end{split}
\end{align}
up to the affine transformation.
Here, $a_i,\, b_i \in \mathbb{Z}_{\geq 0}$ and the notation $\pm a_i$ means that both signs are present.
We call this case a symmetric solution.
We now return to the anomaly cancellation conditions. Under the symmetric constraint, the problem simplifies since Eqs.~(\ref{eq: U(1)X-g-g}) and (\ref{eq: U(1)X3}) are automatically satisfied.
For the ideal case with $n=4$, the remaining condition reduces to finding integer pairs $\{a_1,a_2\}$ and $\{b_1,b_2\}$ such that $a_1^2+a_2^2=b_1^2+b_2^2$.
By contrast, asymmetric solutions must satisfy the full set of original constraints. We will see shortly that ideal solutions to the degree $k=3$ PTE problem are predominantly symmetric.

Some solutions to the degree $k=3$ PTE problem are shown in Table \ref{tab:PTE_k3_ideal_examples}.
Here, the charges are set to be non-negative integers using the above-mentioned affine transformation.
Note that most of them (except for No.10) are symmetric solutions.

Since the gauge coupling constant $\alpha_{\rm X} = g_{\rm X}^2/4\pi$ depends on the energy scale $\mu$ according to the one-loop renormalization group equation
\begin{align}
    \frac{d\alpha_{\rm X}}{d\ln\mu} = \frac{\alpha_{\rm X}^2}{2\pi} \left[ \frac{2}{3} \sum_i q_{{\rm X},i}^2 + \frac{1}{3} \right],
\end{align}
there is an upper bound on the low-energy gauge coupling to avoid a Landau pole 
below the cutoff scale $M_c$.
The solution with the smallest sum of squared charges is $A=\{-3,0,1,4\}$ and $B=\{-2,-2,3,3\}$, for which $\sum q_{{\rm X},i}^2 = 52$. This is also a symmetric solution.
In this case, the coupling constant at the mCP mass scale must satisfy $\alpha_{\rm X} \lesssim 0.006$  in order to avoid a Landau pole 
below the GUT scale $M_c \sim 10^{16}$ GeV.

\begin{table}[t]
\centering
\small
\setlength{\tabcolsep}{4pt}
\begin{tabular}{c|l|l}
No. & $A$ & $B$ \\
\hline
1  & $\{0,4,7,11\}$   & $\{1,2,9,10\}$ \\
2  & $\{0,6,7,13\}$   & $\{1,3,10,12\}$ \\
3  & $\{0,5,10,15\}$  & $\{1,3,12,14\}$ \\
4  & $\{0,7,9,16\}$   & $\{1,4,12,15\}$ \\
5  & $\{0,8,9,17\}$   & $\{2,3,14,15\}$ \\
6  & $\{0,7,11,18\}$  & $\{2,3,15,16\}$ \\
7  & $\{0,6,13,19\}$  & $\{1,4,15,18\}$ \\
8  & $\{0,8,11,19\}$  & $\{1,5,14,18\}$ \\
9  & $\{0,10,11,21\}$ & $\{1,6,15,20\}$ \\
10 & $\{0,11,13,22\}$ & $\{1,7,18,20\}$ \\
\hline
\end{tabular}
\caption{Examples of degree $k=3$ PTE solutions with $n=4$.
}
\label{tab:PTE_k3_ideal_examples}
\end{table}

{\bf Mass Spectrum.--}
The $n\times n$ mass matrix $M$ linking $\psi_i$ with $i=1,\dots,n$ and $\psi_j$ with $j=n+1,\dots,2n$
 is given by
\begin{align}
    M_{ij} \simeq M_c \epsilon^{|a_i - b_j|},
\end{align}
where we have dropped ${\cal O}(1)$ coefficients, and note that the other mass terms are forbidden by the $U(1)_{\rm H}$ symmetry.
Thus, the mass spectrum of the fermions is determined by the charge assignment.

For $k = 3$  we have a lower bound on the size $n$, namely $n \geq k+1 = 4$.
Therefore, one of the robust predictions of our scenario is that there should be at least four mCP mass eigenstates to satisfy the anomaly cancellation conditions. 

For ideal solutions with $k=3$, the solutions are dominated by the symmetric solutions up to the affine transformation.
The symmetric solutions naturally organize into two doublets with comparable mass scales unless accidental cancellation occurs. See Appendix~\ref{sec:two-doublets-proof} for the proof.
This leads to another sharp prediction: if an mCP is discovered, an additional mCP with the same minicharge and a nearby mass is expected.

To make this more explicit, we performed an exhaustive numerical scan over integer charge assignments with $|a_i|,|b_i|\le 22$.
We find $1589$ solutions that satisfy the anomaly cancellation conditions.

\begin{figure*}[!t]
\begin{center}
    \includegraphics[width=0.98\textwidth]{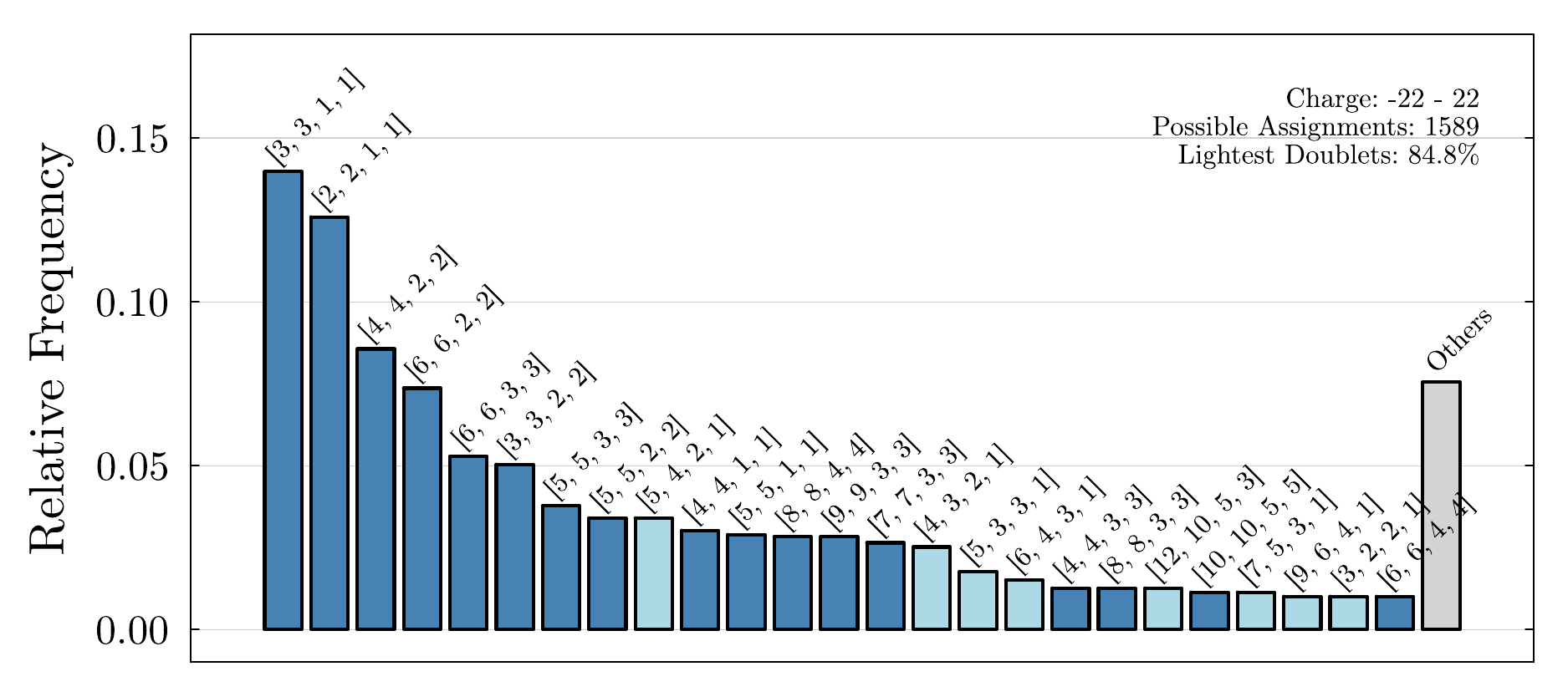}
\end{center}
\caption{%
    Distribution of exponents in the mass hierarchy evaluated by solving a minimum-weight matching problem.
    The frequencies are normalized to unity.
    Bar colors distinguish solutions with a two-doublet spectrum, $e_1=e_2$ and $e_3=e_4$, which correspond to symmetric PTE solutions, from the non-symmetric solutions.
}
\label{fig: exponent_ideal}
\end{figure*}

Fig.~\ref{fig: exponent_ideal} shows the distribution of the mass hierarchy exponents defined by $m_i \sim M_c \epsilon^{e_i}$, where $m_i$ ($i=1,\dots,4$) denotes a mass eigenvalue.
Larger $e_i$ corresponds to a stronger suppression and hence a lighter state.
For each charge assignment we determine $e_i$ by solving a minimum-weight matching problem \cite{Kuhn1955TheHM} on the complete bipartite graph $K_{r,r}\,(r \leq 4)$ with weights $e_{ij} = |a_i - b_j|$.
The matching procedure is described in Appendix~\ref{sec:min-weight-matching}.
We sort $e_i$  in descending order, $e_1 \geq e_2 \geq e_3 \geq e_4$.
The bar color indicates whether the spectrum forms two doublets. We use a dark shade for solutions with $e_1=e_2$ and $e_3=e_4$, and a light shade otherwise. As shown in Appendix~\ref{sec:two-doublets-proof}, symmetric solutions always yield two doublets, so the two-doublet spectra correspond to symmetric PTE solutions. The histogram shows that about $85\%$ of the solutions exhibit a two-doublet structure.

\begin{figure}[!t]
\begin{center}
    \includegraphics[width=0.49\textwidth]{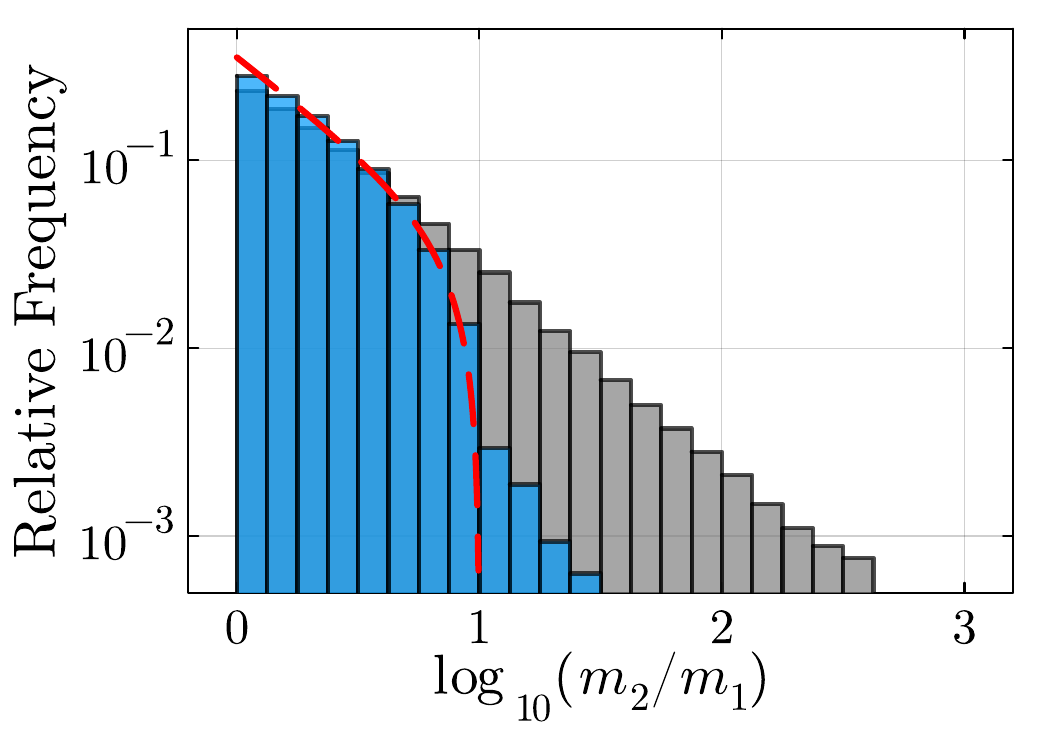}
\end{center}
\caption{%
Mass splitting between the two lightest eigenstates for degree $k=3$ PTE solutions with $n=4$. The grey and blue histograms show the distribution of $m_2/m_1$ with $m_2>m_1$ for coefficient magnitudes sampled in $[0,1]$ and $[0.1,1]$, respectively. The red dashed line shows the estimate based on the diagonal approximation for coefficients with magnitudes in $[0.1,1]$.
}
\label{fig: split}
\end{figure}

Fig.~\ref{fig: split} shows the mass ratio $m_2/m_1$ within the lightest doublet obtained by numerical singular value decomposition of the mass matrix. For each charge assignment, we generate 100 random realizations of the complex $\mathcal{O}(1)$ coefficients, taking their magnitudes to be uniformly distributed in either $[0,1]$ or $[0.1,1]$ and their phases in $[0,2\pi)$, while fixing $\epsilon = 10^{-3}$. The histogram, therefore, quantifies the typical mass splitting of the lightest pair. In most realizations, the two lightest states remain relatively close in mass.

To understand the origin of this distribution, let us consider a simple estimate based on random $\mathcal{O}(1)$ coefficients. The mass splitting in a doublet is expected to be controlled by the ratio of two independent $\mathcal{O}(1)$ coefficients multiplying the same power of $\epsilon$. This corresponds to a diagonal approximation in which mixing effects are neglected. This leads to the probability distribution of $m_2/m_1$,
\begin{multline}
P(z)=\int_X\int_X P_X(x)\,P_X(y)\,
\delta\!\left(z-\max\!\left(\frac{x}{y},\frac{y}{x}\right)\right)\,dx\,dy,
\end{multline}
where $X$ denotes the range of absolute values of the random coefficients and $P_X$ is their distribution. For the red dashed line in Fig.~\ref{fig: split}, we take $X=[0.1,1]$. As seen in the figure, this estimate captures the histogram well for modest splittings. The small tail extending beyond the red dashed line arises from mixing effects neglected in the diagonal approximation. In particular, when mixing induces an accidental cancellation in $\det M$, the lightest eigenvalue can become exceptionally small, resulting in a larger mass ratio $m_2/m_1$.

{\bf Conclusion.--}
We established an equivalence between anomaly-free chiral minicharged sectors and the degree $k=3$ Prouhet-Tarry-Escott problem.
This correspondence implies a minimal multiplicity of four mCP mass eigenstates and, for minimal ideal solutions, it typically yields a near-degenerate doublet structure.

For mCP models without an unbroken $U(1)_{\rm H}$, our framework does not apply directly because the PTE equivalence relies on mixed anomaly cancellation conditions, and one should instead use the standard anomaly constraints for a single $U(1)$~\cite{Nakayama:2011dj,Costa:2019zzy}.

In our setup, spontaneous breaking of $U(1)_{\rm X}$ implies a massive gauge boson $X$.
If $X$ kinetically mixes with the SM hypercharge, it can be produced in collider and fixed-target experiments and decay into mCP pairs~\cite{Izaguirre:2015eya}.
While we have kept the identity of $U(1)_{\rm X}$ generic, a well-motivated choice is $U(1)_{B-L}$, to which our framework applies straightforwardly.
Although gauged $U(1)_{B-L}$ requires three right-handed neutrinos for anomaly cancellation, they are $U(1)_{\rm H}$ singlets and hence do not mix with the mCP sector.
Additional chiral fermions charged under $U(1)_{B-L}$ may also render the lightest state a dark-matter candidate~\cite{Nakayama:2011dj,Nakayama:2018yvj}.
We note that our argument remains valid even in the presence of kinetic mixing between $U(1)_{\rm X}$ and hypercharge.

Probing the predicted near-degenerate doublet structure of mCPs requires experimental sensitivity to the mCP mass spectrum. Determining mCP masses is itself a dedicated experimental challenge. In principle, the mass could be inferred from several observables, such as the energy spectrum of the detected signal, the facility's production energy, and/or time-of-flight measurements, provided the mCPs are produced in a semi-relativistic or non-relativistic regime.

\section*{Acknowledgments}
This work was supported by JSPS KAKENHI 25H02165 [FT] and 25KJ0564 [JL].
This work was also supported by the World Premier International Research Center Initiative (WPI), MEXT, Japan, and is based upon work from COST Action COSMIC WISPers CA21106, supported by COST (European Cooperation in Science and Technology).
One of us (JL) is supported by the Graduate Program on Physics for the Universe (GP-PU), Tohoku University.
YDT is supported by the Dorothy Hodgkin Fellowship, funded by the Royal Society, United Kingdom, and appreciates the start-up support from the University of Manchester.

\bibliographystyle{apsrev4-1}
\bibliography{ref}

\clearpage
\section*{Supplemental Material}
\addcontentsline{toc}{section}{Supplemental Material}

\setcounter{section}{0}
\renewcommand{\thesection}{S\arabic{section}}
\renewcommand{\thesubsection}{S\arabic{section}.\arabic{subsection}}

\section{Proof of symmetric solutions leading to multiplicity in mass spectrum}
\label{sec:two-doublets-proof}

Here, we show that symmetric solutions with even $n=2m$ lead to a pairwise-degenerate mass spectrum. We have focused on the $n=4$ case in the main text, but the argument applies to general symmetric solutions with even $n$ because the mass matrix can be organized into parity-even and parity-odd sectors, as we will see shortly.

We assume that the two charge multisets are centrally symmetric
about a common center. Equivalently, after a common shift (which does not affect charge
differences), we may write the solution in the sign-paired form
\begin{multline}
    A=\{\pm a_1, \cdots , \pm a_m\}, \ 
    B=\{\pm b_1, \cdots , \pm b_m\}, \\
    (a_i>0,\ b_j>0),
\label{eq:sign-paired}
\end{multline}
where the notation $\pm a_i$ means that both $+a_i$ and $-a_i$ are included in the solution.

We assume that the (Dirac) mass matrix has the hierarchical form
\begin{equation}
M_{ab}\;\simeq\; M\,\epsilon^{|a-b|},
\qquad \epsilon\ll 1,
\label{eq:mass-ansatz}
\end{equation}
where $a\in A$ labels rows and $b\in B$ labels columns. Overall $\mathcal{O}(1)$ coefficients
do not affect the leading parametric argument as long as they respect the sign-pairing symmetry
discussed below.

Let $P_A$ be the permutation matrix that exchanges $+a_i\leftrightarrow -a_i$ for each $i$ on the
row space, and similarly $P_B$ exchanges $+b_j\leftrightarrow -b_j$ on the column space.
Using \eqref{eq:mass-ansatz},
\begin{equation}
    (P_A M P_B)_{a b}=M_{-a,-b}
    =M\,\epsilon^{|a-b|}
    =M_{ab},
\end{equation}
so we have the invariance
\begin{equation}
P_A M P_B = M.
\label{eq:Z2-invariance}
\end{equation}

For each $\{a_i,-a_i\}$ pair we introduce the even/odd combinations
\begin{align}
    \begin{split}
        &|a_i\rangle_{\pm}\equiv \frac{1}{\sqrt{2}}\left(|a_i\rangle\pm |-a_i\rangle\right), \\
        &|b_j\rangle_{\pm}\equiv \frac{1}{\sqrt{2}}\left(|b_j\rangle\pm |-b_j\rangle\right).
    \end{split}
\label{eq:even-odd-basis}
\end{align}
They satisfy $P_A|a_i\rangle_{\pm}=\pm|a_i\rangle_{\pm}$ and $P_B|b_j\rangle_{\pm}=\pm|b_j\rangle_{\pm}$.
Then, for opposite parities, we obtain
\begin{align}
    {}_{\pm}\!\langle a_i|\,M\,|b_j\rangle_{\mp}
    &={}_{\pm}\!\langle a_i|\,P_A M P_B\,|b_j\rangle_{\mp}
    \nonumber\\
    &=-\,{}_{\pm}\!\langle a_i|\,M\,|b_j\rangle_{\mp},
\end{align}
which implies ${}_{\pm}\!\langle a_i|\,M\,|b_j\rangle_{\mp}=0$.
Therefore, in the basis \eqref{eq:even-odd-basis} the mass matrix becomes block diagonal,
\begin{equation}
\begin{pmatrix}
M^{(+)} & 0\\
0 & M^{(-)}
\end{pmatrix},
\label{eq:block-diag}
\end{equation}
where $M^{(+)}$ and $M^{(-)}$ are $m\times m$ matrices acting within the even and odd subspaces,
respectively.

Using \eqref{eq:mass-ansatz} and \eqref{eq:even-odd-basis}, one finds
\begin{align}
    \begin{split}
        M^{(+)}_{ij}
        &=\frac{1}{2}\big(M_{a_i b_j}+M_{a_i,-b_j}+M_{-a_i,b_j}+M_{-a_i,-b_j}\big) \\
        &= M\Big(\epsilon^{|a_i-b_j|}+\epsilon^{a_i+b_j}\Big),
    \end{split}
\end{align}
\begin{align}
    \begin{split}
        M^{(-)}_{ij}
        &=\frac{1}{2}\big(M_{a_i b_j}-M_{a_i,-b_j}-M_{-a_i,b_j}+M_{-a_i,-b_j}\big) \\
        &= M\Big(\epsilon^{|a_i-b_j|}-\epsilon^{a_i+b_j}\Big).
    \end{split}
\end{align}
Since $a_i,b_j>0$ implies $a_i+b_j>|a_i-b_j|$, we have
$\epsilon^{a_i+b_j}\ll \epsilon^{|a_i-b_j|}$ for $\epsilon\ll1$, and hence
\begin{equation}
M^{(+)} = M^{(-)} + \mathcal{O}\!\left(M\,\epsilon^{a_i+b_j}\right).
\label{eq:blocks-equal}
\end{equation}
In the strict hierarchical limit $\epsilon\to0$, the two blocks become identical. Thus, symmetric solutions indeed lead to doublets.

\section{Systematic way to find mass hierarchy}
\label{sec:min-weight-matching}

With given charge assignment $\{a_i\}$ and $\{b_j\}$, the mass matrix elements are given by powers of a small parameter $\epsilon$ as
\begin{align}
    M_{ij} = y_{ij} M_c \epsilon^{e_{ij}},
\end{align}
where $e_{ij} = |a_i - b_j|$.
The diagonal mass matrix,
\begin{align}
    D_{ij} = m_i \delta_{ij},
\end{align}
is obtained by a proper transformation $D = U_{\rm L} M U_{\rm R}^\dagger$ with unitary matrices $U_{\rm L}$ and $U_{\rm R}$.
The exponent in the mass hierarchy, $e_i$, is defined by rounding $\tilde{e}_i$ derived from
\begin{align}
    m_i = M_c \epsilon^{\tilde{e}_i}.
\end{align}
Here, all mass eigenvalues $m_i$ are sorted in ascending order, $0<m_1 \leq m_2 \leq \cdots \leq m_n$.

To find the mass hierarchy, we consider the quantities $\Delta_r$ defined by
\begin{align}
    \Delta_r(M) \equiv \sum_{I, J} \left| \det M_{IJ} \right| ^2,
    \label{eq:def_Delta}
\end{align}
where $M_{IJ}$ is a $r \times r$ submatrix of $M$ specified by the row index set $I = \{i_1, \dots, i_r\}$ and the column index set $J = \{j_1, \dots, j_r\}$.
Namely, the sum is taken over all possible choices of $I$ and $J$, in other words, all possible $r \times r$ submatrices of $M$.
Note that $\Delta_r (M)$ remains unchanged under the unitary transformations of $M$, $\Delta_r (U_{\rm L} M U_{\rm R}^\dagger) = \Delta_r (M)$ for any unitary matrices $U_{\rm L}$ and $U_{\rm R}$, and thus can be expressed in terms of the mass eigenvalues.
In fact, it can be shown that
\begin{align}
    \Delta_r (M) = \sum_{J} m_{j_1}^2 \cdots m_{j_r}^2,
    \label{eq:rel_Delta_mass}
\end{align}
where $m_i$ are mass eigenvalues, the singular values of $M$.
The proof is as follows:
The generalized Cauchy-Binet formula gives
\begin{align}
    \det[(M M^\dagger)_{II}] = \sum_{J} \left| \det [M_{IJ}] \right|^2.
\end{align}
Now the sum over $I$ leads to
\begin{align}
     \sum_{I} \det[(M M^\dagger)_{II}] = \sum_{IJ} \left| \det [M_{IJ}] \right|^2 = \Delta_r (M).
\end{align}
The relation between the left-hand side and mass eigenvalues is given by matching the coefficients of the characteristic polynomial of $M M^\dagger$.
The characteristic polynomial $p(\lambda)$ of $M M^\dagger$ is given by
\begin{align}
    p(\lambda) &= \det (\lambda I - M M^\dagger) \\
    &= \lambda^n - \left( \sum_i m_i^2 \right) \lambda^{n-1} + \left( \sum_{i<j} m_i^2 m_j^2 \right) \lambda^{n-2} - \cdots \nonumber \\
    &\quad + (-1)^n (m_1^2 m_2^2 \cdots m_n^2).
\end{align}
On the other hand, by expanding the determinant, we have
\begin{align}
    p(\lambda) &= \sum_{r=0}^{n} (-1)^r \left( \sum_{I} \det[(M M^\dagger)_{II}] \right) \lambda^{n-r} \\
    &= \sum_{r=0}^{n} (-1)^r \Delta_r (M) \lambda^{n-r},
\end{align}
where $\Delta_0 = 1$.
By comparing the two expressions, we obtain Eq.~(\ref{eq:rel_Delta_mass}).

Assuming a sufficient hierarchy among mass eigenvalues, $\epsilon \ll 1$, the leading contribution to $\Delta_r (M)$ comes from the heaviest $r$ mass eigenvalues as
\begin{align}
    \Delta_r (M) \simeq m_{n}^2\, m_{n-1}^2 \cdots m_{n-r+1}^2,
\end{align}
assuming non-vanishing mass eigenvalues.
Thus, by evaluating $\Delta_r (M)$ for $r = 1, \dots, n$, we can find the fermion masses as
\begin{align}
    m_n^2 &\simeq \Delta_1 (M), \\
    m_{n-1}^2 &\simeq \frac{\Delta_2 (M)}{\Delta_1 (M)}, \\
    &\vdots \nonumber \\
    m_1^2 &\simeq \frac{\Delta_n (M)}{\Delta_{n-1} (M)}.
\end{align}
Here, we are interested in the exponents in the mass hierarchy.
Let us define the exponent $E_r$ by rounding $\tilde{E}_r$ derived from
\begin{align}
    \Delta_r (M) = M_c^{2r} \epsilon^{2 \tilde{E}_r}.
    \label{eq:def_Ek}
\end{align}
Then, the exponents in the mass hierarchy are given by
\begin{align}
    e_n &= E_1, \\
    e_{n-1} &= E_2 - E_1, \\
    &\vdots \nonumber \\
    e_1 &= E_n - E_{n-1}.
\end{align}
Connecting Eqs.~(\ref{eq:def_Delta}) and (\ref{eq:def_Ek}), we can obtain
\begin{align}
    \epsilon^{2\tilde{E}_r}
    = \sum_{I,J} \left| \sum_{\pi \in S_r} {\rm sgn}(\pi) \prod_{q=1}^{r} \epsilon^{e_{i_q j_{\pi (q)}}} \right|^2
\end{align}
where $S_r$ is the permutation group of order $r$.
The leading contribution to the right-hand side comes from the choice of $I$, $J$, and $\pi$ that minimizes the exponent $\sum_{q=1}^{r} e_{i_q j_{\pi (q)}}$, and thus
\begin{align}
    E_r = \min_{I,J} \left[ \min_{\pi \in S_r} \left[\sum_{q=1}^{r} e_{i_q j_{\pi (q)}}\right] \right].
\end{align}

This minimization problem is equivalent to the minimum-weight matching problem on a complete bipartite graph $K_{r,r}$ with two disjoint vertex sets $U = \{i_1, \dots, i_r\}$ and $V = \{j_1, \dots, j_r\}$.
We assign weights $e_{pq} = |a_p - b_q|$ to the edges connecting $i_p$ and $j_q$.
A perfect matching is a subset of edges such that every vertex is connected to exactly one edge in the subset.
The weight of a perfect matching is defined as the sum of the weights of the edges in the matching.
The minimum-weight matching problem aims to find a perfect matching with the smallest weight.
In our case, the minimum-weight matching corresponds to the leading order contribution to $E_r$.
The minimum-weight matching problem can be solved systematically by the Hungarian algorithm \cite{Kuhn1955TheHM}.

\section{Numerical Analysis in the Ideal Case}
\label{sec: numerical ideal case}

\begin{figure*}[!t]
\begin{center}
    \includegraphics[width=0.98\textwidth]{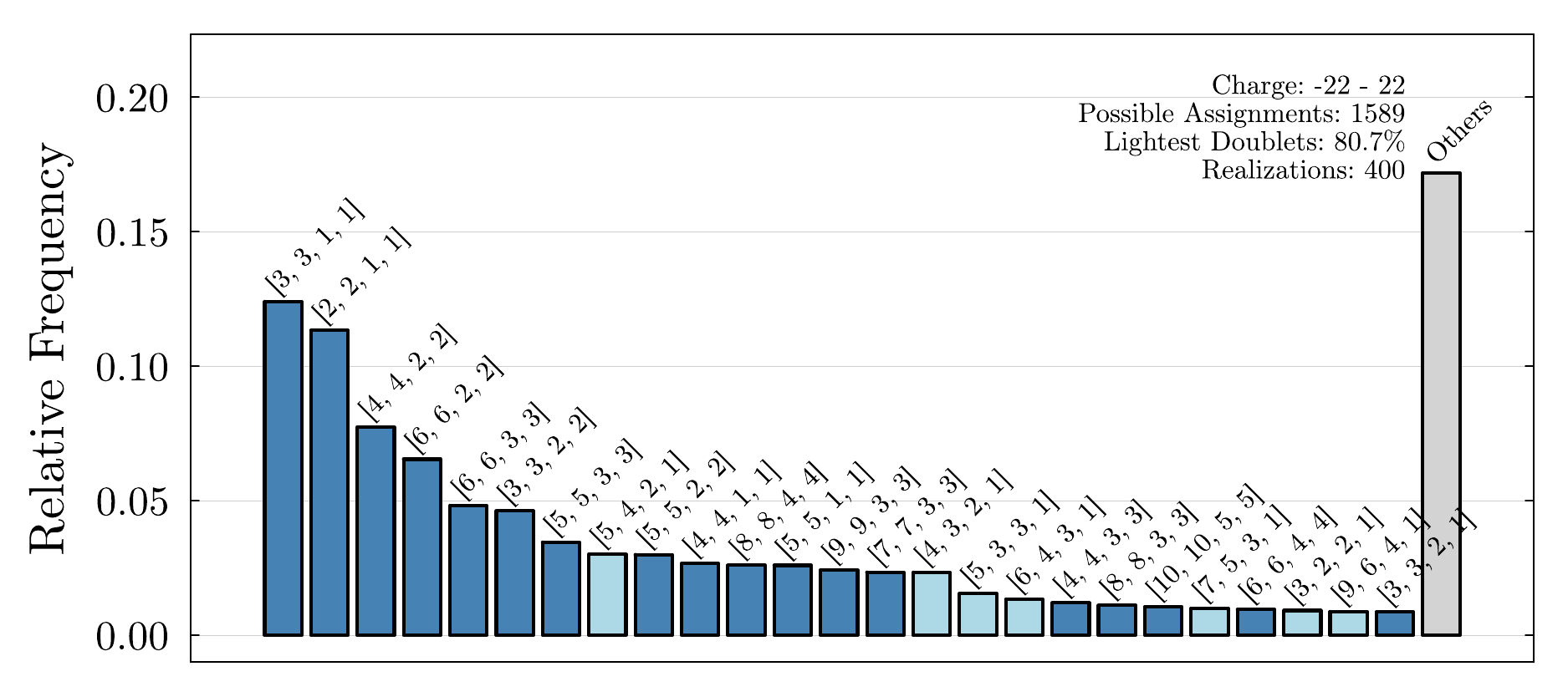}
\end{center}
\caption{%
    Mass eigenvalue spectrum obtained by numerical singular value decomposition of the mass matrix.
    Random $\mathcal{O}(1)$ coefficients $y_{ij}$ are generated for each realization and we set $\epsilon = 10^{-3}$.
    The result is averaged over 400 realizations for each charge assignment.
    The bar color distinguishes solutions with a light-doublet spectrum, $e_1=e_2$.
    We confirmed that the result approaches the expectation from the minimum-weight matching problem as $\epsilon$ becomes smaller.
}
\label{fig: exponent_ideal_num}
\end{figure*}

In the main text, we have shown the mass hierarchy exponents $e_i$ obtained by solving the minimum-weight matching problem.
On the other hand, we can also obtain the mass eigenvalues by numerical singular value decomposition of the mass matrix to address the effect of the random coefficients and insufficient hierarchy.
The exponent $e_i$ of each mass eigenvalue $m_i$ is calculated by rounding $\tilde{e}_i$ derived by $m_i = M_c \epsilon^{\tilde{e}_i}$.
Fig.~\ref{fig: exponent_ideal_num} shows the mass eigenvalue spectrum obtained by numerical singular value decomposition of the mass matrix.
All combinations of $e_i$ are sorted in descending order, $e_1 \geq e_2 \geq e_3 \geq e_4$.
The bar color indicates whether the spectrum forms light doublets. We use a dark shade for solutions with $e_1=e_2$, and a light shade otherwise.
Using the benchmark value $\epsilon = 10^{-3}$, we generated random $\mathcal{O}(1)$ coefficients $y_{ij}$ to make specific fermion mass matrix for each realization.
In fact, the choice of measure for $|y_{ij}|$ is arbitrary and depends on the UV completion of the model, whereas the measure of $\arg(y_{ij})$ is fixed by the Haar measure of $U(1) \times U(1)$, which is effectively uniform, as for $U(1)$~\cite{Haba:2000be}.
For the analysis in this section, we generate random coefficients with absolute values uniformly distributed in $[0,1]$ and phases uniformly distributed in $[0,2\pi)$. In Fig.~\ref{fig: split} of the main text, we compare this choice with $[0.1,1]$.
The result shows differences in detail from Fig.~\ref{fig: exponent_ideal} because of the cancellation and mixing effects, reducing the fraction of solutions with a light doublet to about $81\%$, but the overall feature is consistent with the result obtained by solving the minimum-weight matching problem.
We confirmed that the distribution of $e_i$ approaches the one in Fig.~\ref{fig: exponent_ideal} as $\epsilon$ becomes smaller.

\section{Non-Ideal Cases}
\label{sec: non-ideal case}

\begin{figure*}[!t]
\begin{center}
    \includegraphics[width=0.98\textwidth]{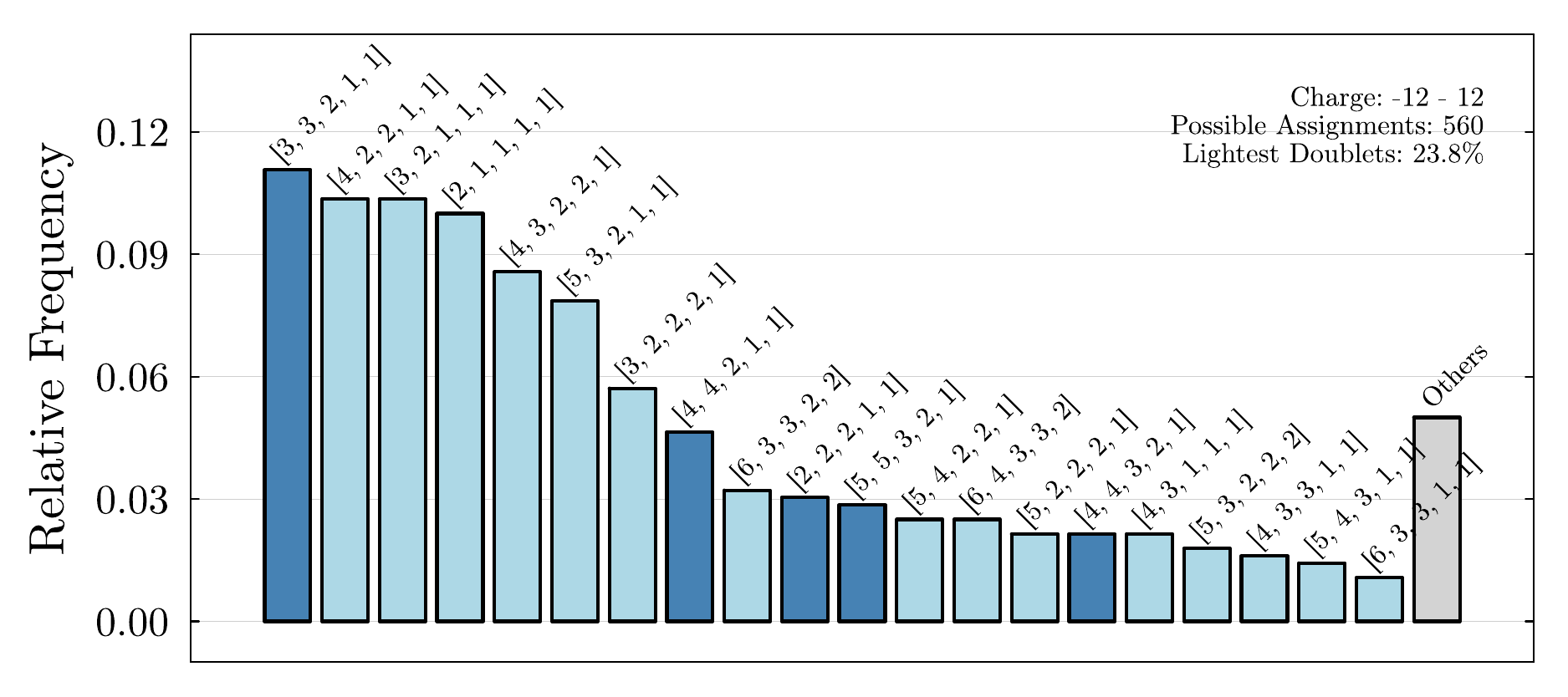}
\end{center}
\caption{%
    The same as Fig.~\ref{fig: exponent_ideal} but for $n=5$.
    The bar color distinguishes solutions with a light-doublet spectrum, $e_1=e_2$.
    Only about $24\%$ of the solutions exhibit a lightest doublet, which is much smaller than the case of $n=4$.
}
\label{fig: exponent_n5}
\end{figure*}

Here we consider the case of $n=5$, which is the minimal size for non-ideal solutions.
For odd $n$, we can generalize the symmetric solution as
\begin{align}
    \begin{split}
        A&=\{c_1,\ldots, c_n\},\\
        B&=\{-c_n,\ldots, -c_1\},
    \end{split}
\end{align}
up to an affine transformation where $c_i \in \mathbb{Z}$ satisfy $\sum_i c_i = \sum_i c_i^3 = 0$.
In fact, this case is rare, and we find only 12 solutions among 560 solutions obtained by scanning over integer charge assignments with $|a_i|,|b_i|\le 12$.
The results of the mass hierarchy analysis for $n=5$ are shown in Fig.~\ref{fig: exponent_n5}.
We use a dark bar color for solutions with $e_1=e_2$, indicating the presence of a lightest doublet, and a light shade otherwise.
Only about $24\%$ of the solutions exhibit a lightest doublet, which is much smaller than the case of $n=4$.
The multiplicity of the lightest state is not a generic feature for odd $n$.

\begin{figure*}[!t]
\begin{center}
    \includegraphics[width=0.98\textwidth]{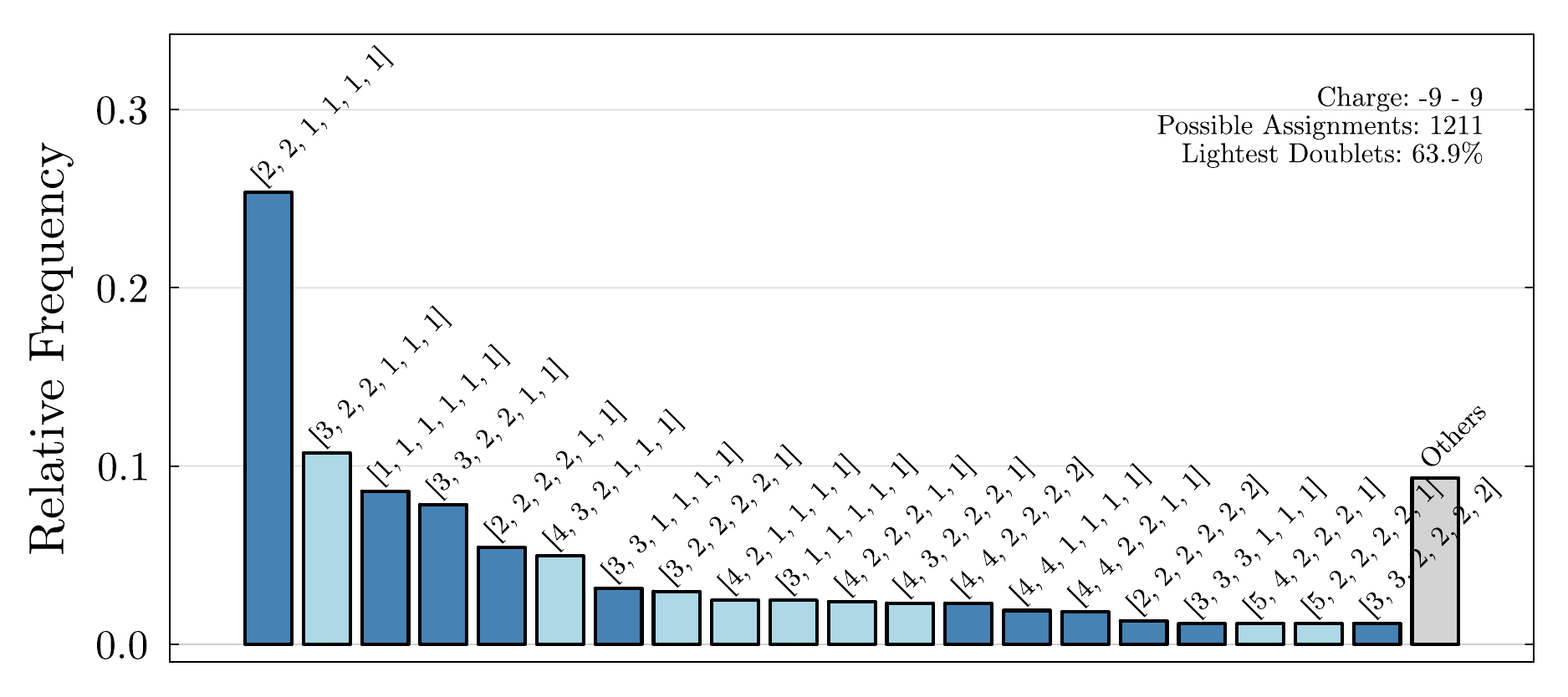}
\end{center}
\caption{%
    The same as Fig.~\ref{fig: exponent_ideal} but for $n=6$.
    The bar color distinguishes solutions with a light-doublet spectrum, $e_1=e_2$.
    About $64\%$ of the solutions exhibit a lightest doublet.
}
\label{fig: exponent_n6}
\end{figure*}

\begin{figure*}[!t]
\begin{center}
    \includegraphics[width=0.98\textwidth]{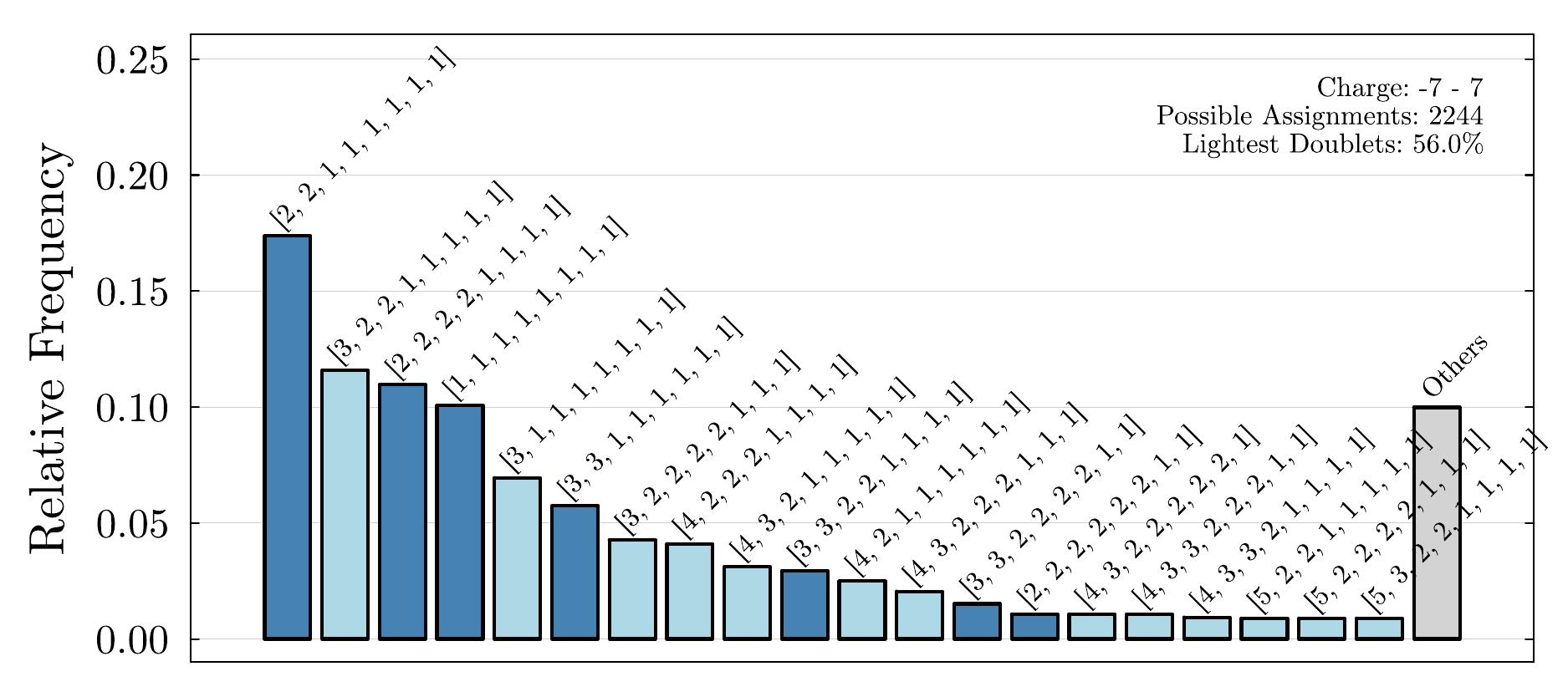}
\end{center}
\caption{%
    The same as Fig.~\ref{fig: exponent_ideal} but for $n=8$.
    The bar color distinguishes solutions with a light-doublet spectrum, $e_1=e_2$.
    About $56\%$ of the solutions exhibit a lightest doublet.
}
\label{fig: exponent_n8}
\end{figure*}

We can check whether the multiplicity is the generic feature for even $n$ cases by analyzing the case of $n=6$.
We find 1211 solutions for $n=6$ by scanning over integer charge assignments with $|a_i|,|b_i|\le 9$.
The results of the mass hierarchy analysis for $n=6$ are shown in Fig.~\ref{fig: exponent_n6}.
About $64\%$ of the solutions exhibit a lightest doublet, which is a majority but smaller than the case of $n=4$.
We also checked the case of $n=8$ (Fig.~\ref{fig: exponent_n8}) and found that about $56\%$ of the solutions exhibit a lightest doublet.
The notable finding is that the multiplicity feature is most pronounced in the ideal solution case ($n=4$), which is also favored from the perspective of Occam's razor and the Landau pole constraint.
Additionally, among the most frequent mass hierarchies across all solutions, the ideal case $n=4$ predominantly exhibits $e_1 = e_2 = 3$, in contrast to the $n > 5$ cases, which favor $e_1 = e_2 = 2$.
This result is accidental, but it is interesting that the ideal solution case favors a lighter doublet than the non-ideal cases.

\end{document}